\documentclass[11pt,draft]{article}
 \usepackage{amsfonts}
 \usepackage{amssymb}
 \newfont{\bbbold}{msbm10}

 \def\bbC{\mbox{\bbbold C}}

 \def\bbH{\mbox{\bbbold H}}

 \def\bbM{\mbox{\bbbold M}}

 \def\bbP{\mbox{\bbbold P}}
 
 \def\bbR{\mbox{\bbbold R}}

 \def\cA{{\cal A}}

 \def\cD{{\cal D}}
 
 \def\cF{{\cal F}}
 \def\cG{{\cal G}}
 \def\cH{{\cal H}}

 \def\cL{{\cal L}}
 \def\cM{{\cal M}}
 \def\cN{{\cal N}}
 \def\cO{{\cal O}}

 \def\cT{{\cal T}}

 \newfont{\goth}{eufm10 scaled \magstep1}

 \def\gg{\mbox{\goth g}}
 
 \def\gi{\mbox{\goth i}}

 \def\go{\mbox{\goth o}}

 \def\gs{\mbox{\goth s}}

 \def\a{\alpha}
 \def\b{\beta}
 \def\C{\Gamma}
 \def\d{\delta}
 \def\e{\epsilon}
 
 \def\z{\zeta}

 \def\l{\lambda}\def\L{\Lambda}

 \def\r{\rho}
 \def\s{\sigma}
 
 \def\th{\theta}

 \def\be{\begin{equation}}\def\ee{\end{equation}}
 \def\bea{\begin{eqnarray}}\def\eea{\end{eqnarray}}
 \def\ba{\begin{array}}\def\ea{\end{array}}
 
 \def\o{\omega}\def\O{\Omega}
 \def\del{\partial}

 \def\nno{\nonumber}
 \def\del{\partial}

 \let\la=\label

 {}

 \def\bd{\begin{document}}
 \def\ed{\end{document}}
 \def\bea{\begin{eqnarray}}\def\barr{\begin{array}}\def\earr{\end{array}}
 \def\eea{\end{eqnarray}}
 \def\ft#1#2{{\textstyle{{\scriptstyle #1}\over {\scriptstyle #2}}}}
 \def\fft#1#2{{#1 \over #2}}
 \newcommand{\eq}[1]{(\ref{#1})}
 \def\eqs#1#2{(\ref{#1}-\ref{#2})}
 \def\det{{\rm det\,}}
 \def\tr{{\rm tr}}\def\Tr{{\rm Tr}}

\begin{document}

 \thispagestyle{empty}

 \vspace{20pt}

 \begin{center}
 {\Large{\bf Supersymmetry, a Biased Review}}\footnote{Invited Lectures at ``The
22nd Winter School GEOMETRY AND PHYSICS, Srni, Czech Republic, January 12-19,
2002}
 \vspace{30pt}

{U. Lindstr\"om}\footnote{e-mail: ul@physto.se} \vskip
.2cm{Department of Physics}, \linebreak
{Stockholm University},
SCFAB\\
{S-106 91 Stockholm, Sweden}\\
\vspace {15pt}

 \vspace{90pt}

 \end{center}

{\bf Abstract}\\

This set of lectures contain a brief review of some basic supersymmetry and its
representations, with emphasis on superspace and superfields. Starting from the
Poincar\'e group, the supersymmetric extensions allowed by the
Coleman-Mandula theorem and its generalisation to superalgebras, the
Haag, Lopuszanski and Sohnius theorem, are discussed. Minkowski space is
introduced as a quotient space and Superspace is presented as a direct
generalization of this. The focus is then shifted from a general presentation to
the relation between supersymmetry and complex geometry as manifested in the
possible target space geometries for $N=1$ and $N=2$ supersymmetric nonlinear
sigma models in four dimensions. Gauging of isometries in nonlinear sigma
models is discussed for these cases, and the quotient construction is described.
\vspace{8cm}
\pagebreak

\tableofcontents

\section{Introduction}
In these lectures I try to give a physicists picture of (some
aspects of) supersymmetry and its representations. Since the
majority of the audience at the meeting were mathematicians, I
presented a lot of background that is normally taken for
granted. In taking this course, the choice
of what to include and what to leave out becomes even more
difficult than is usually the case. Ideally, had I been able to request a similar
contribution from a mathematician I myself would have wanted a translation table
of the kind ``When they say... they mean...'', but the present article is no such
thing. 
 In the end, in the
written version, I have included additional explanations wehere I feel that a few
words may clarify the presentation substantially for a newcomer to the subject. 
One problem is, of course, that I don't know what points may create difficulties.
I don't want to make too much of the ``clash of cultures'', but it is
abundantly clear that, e.g., the use of indices creates one
communication problem. Rather than modifying the presentation
to conform with an index-free notation, however, I have kept to
the physicists notation in a hope that a reader may use the
text to understand how indices are used to keep track of
transformation properties in the physics literature.

I must stress that the title I have given my contribution is correct; I have
only included the material I thought I needed to get to the relation between
supersymmetry and complex geometry as soon as possible without totally sacrifying
the general picture. This is also reflected in my list of references, which is
sadly inadequate. However, I believe it includes enough standard
texts (\cite{Gates:1983nr}, \cite{Bagger:1990qh},
\cite{Freund:1986ws},
\cite{Buchbinder:1998qv}) that the reader may find his way to all the basic
sources through them. A couple of the general references are further particularly
suited for a mathematical audience, namely \cite{Freed:1999mn} and
\cite{Deligne:1999qp}.

The lectures are divided into two parts, introductory material (Sections 1, 2 and
3) and the relation between supersymmetry and complex geometry.. (Section 4). For
the first part I draw from numerous sources, consiously and subconsiouly. For the
second part my main material is the three articles \cite{Lindstrom:1983rt},
\cite{Hull:1986pq} and \cite{Hitchin:1987ea}.

\section{Relativistic symmetries}

References for this section are the text books and articles referred to in the
introduction along with any good book on quantum field theory such as
\cite{Peskin:1995ev} or
\cite{Weinberg:1995mt}. Also the (old) review articles \cite{Fayet:1977cr} and
\cite{Salam:1978ib} may provide useful background. If one is more generally
interested in graded algebras and their representations  samples of the possible
references are \cite{Corwin:1975fi}, \cite{Pais:1975hu},
\cite{Rittenberg:1978eg},
\cite{Scheunert:1977wj},
\cite{Scheunert:1977wi},
\cite{Scheunert:1976ug},
\cite{Scheunert:1976uf},
\cite{Nahm:1976mu}. 

\subsection{The Poincar\'e algebra}
In theoretical high-energy physics we study the motion of
particles, strings and branes in various ambient space times.
This means that we are interested in manifolds with a range of
dimensions, from zero spacelike and one timelike (the
particle) to 25 spacelike and one timelike (the target
space\footnote{Target space and ambient space-time are synonymous
in this text.} of the bosonic string). In all these dimensions
we mainly focus on relativistically invariant models, however.
Thus the fundamental structure is given by the (tangent space)
group
$ISO(D-1,1)$, the $D$-dimensional Poincar\'e group. The
generators of its Lie-algebra $\gi\gs\go(D-1,1)$ satisfy the
following algebra
\bea
&&[P_a,P_b]=0\nno\\
&&[M_{ab},P_c]=\frac{i}{2}\eta_{c[a}P_{b]}\nno\\
&&[M_{ab},M_{cd}]=\frac{i}{2}\eta_{c[a}M_{b]d}-c\leftrightarrow
d~,
\la{poin}
\eea
where\footnote{We use the bracket notation to indicate symmetry or
skewness, i.e. (ab) denotes symmetrization and [ab] denotes
antisymmetrization, with no combinatoical factors.}
$P_a$ generate translations,
$M_{ab}$ generate Lorentz transformations, and $\eta$ is the 
Minkowski metric whose
relation to the spacetime metric $g$ is 
\be
ds^2=g_{mn}dX^mdX^n=\eta_{ab}e^a_me^b_ndX^mdX^n~,
\ee
where the line-element $ds^2$ is expressed in using the
coordinates $X^m$ in the space-time. The one forms
$e^a=e^a_mdX^m$ are often called ``viel-beins'' in the physics
literature. As a further note on notation, ``curved''
(space-time) indices are taken from the middle of the alphabet,
tangent space indices are from the beginning of the alphabet
and the summation convention is used (repeated indices are
summed over). 

The algebra (\ref{poin}) shows that $P_a$ transforms in the
fundamental representation of the Lorentz group (LG), i.e., as a
vector, and that $M_{ab}$ itself transforms as an
(antisymmetric) second rank tensor.

One is typically interested only in the \underline{proper}
Lorentz group, $SO(D-1,1)^\uparrow$, given by matrices $\L~:
\L^T\eta\L=\eta,~det\L=+1,~\L^0_0\geq 0$. This semi-simple
Lie-group is not simply connected. Its universal covering group
is $Spin(D-1,1)$. An element in the fundamental representation
of this group is called a spinor and we will denote it by
$\Psi_\a\in T_S$. All (finite dimensional) representations of
the LG may be obtained from tensor products
$T_S\otimes T_S\otimes T_S...$, a useful fact that, e.g., later
allows us to use pairs of spinor indices to represent
vector indices.

The representations of the LG fall into two distinct classes,
those with integer spin, the \underline{bosons}, and those with half integer
spin, the \underline{fermions}. To be more precise, the names refer to the
elementary particles that transform in the corresponding
representations. Fermions are the constituents of matter and
bosons govern the forces in nature. The obey different
statistics; many bosons can occupy the same state (c.f.
Bose-Einstein condensate) while only one fermion can be in a
particular state in the Hilbert space (the Pauli exclusion principle).

\subsection{Minkowski space $\cM$}

A useful way of representing  the Poincar\'e
group  is in terms of fields\footnote{I am using the
word ``field'' in the standard physicist way, meaning
(usually $C^\infty$) functions.} over Minkowski space $\cM$,
with the generators of the algebra (\ref{poin}) described by
differential operators acting on these fields. The Minkowski
space itself may be thought of as the quotient of the
Poincar\'e group  with the Lorentz group
\be
ISO(D-1,1)/SO(D-1,1)~.
\la{mink}
\ee
Since this is analogous to the way in which Superspace is
defined in subsection 3.1, it pays to look at the construction in some
detail in this simpler context.\\

A point in $\cM$ is parametrized as 
\be
h(x)=e^{ix^aP_a}\bf{1}~,
\ee
and the group acts by left multiplication
\be
h(gx)=h(x')\equiv gh(x) modSO(D-1,1)~.
\ee
For a translation  $g=e^{i\xi^aP_a}$, with parameter $\xi$, this
yields
\bea
gh&=&e^{i\xi^aP_a}e^{ix^aP_a}=e^{i(x^a+\xi^a)P_a+\frac{i}{2}[\xi
P,xP]+...}\nno\\
&&=e^{i(x^a+\xi^a)P_a}\nno\\
&&\equiv h(x')~,
\la{tran}
\eea
where the Baker-Campbell-Hausdorff (BCH) formula becomes trivial
since translations commute, as seen in (\ref{poin}). A
translation thus induces the following coordinate change:
\be
x'^a=x^a+\xi^a, \quad \Rightarrow \delta x^a=\xi^a ~,
\la{itran}
\ee
where the last relation gives the infinitesimal transformation.
The corresponding calculation for a Lorentz transformation
$g=e^{i\omega^{ab}M_{ab}}$ with parameter $\omega^{ab}$ is less
trivial since the generators $P$ and $M$ don not commute. The BCH
formula thus contributes nontrivially. One also has to make use
of the quotient structure when calculating
\bea
h(x')&=&gh=e^{i\omega^{ab}M_{ab}}e^{ix^cP_c}\nno\\
&&=e^{i\omega^{ab}M_{ab}}e^{ix^cP_c}e^{-i\omega^{ab}M_{ab}}\nno\\
&&=e^{x^c(e^{i\omega M}P_ce^{-i\omega M})}\nno\\
&&=e^{x^c(e^{\omega})_c^aP_a}~,
\eea
where all operations are performed $modSO(D-1,1)$. The induced action of
on the coordinate is thus
\be
x'^a= x^b(e^{\omega})_b^a\quad \Rightarrow \delta x^a=x^b\omega_b^a ~.
\la{lore}
\ee

\subsection{Fields over $\cM$}

Now that we know howe the Poincar\'e transformations act on the Minkowski
coordinate $x$, we find representations in terms of (scalar) fields $f$ by
requiring that they transform as
\be
f'(x')=f(x)~.
\ee
Under an infinitesimal transformation $x\to x+\delta x$, the fields thus
obey
\be
\delta f(x)\equiv f'(x)-f(x) = -\delta x^a\partial _a f(x)~,
\la{delta}
\ee
where $\partial_a \equiv \partial /\partial x^a$, and (\ref{delta})
defines what we are to mean by the infinitesimal variation of a field.
Inserting the infinitesimal coordinate transformations in (\ref{itran}) and
(\ref{lore}), we find the action of  a translation or a Lorentz
transformation on a scalar field. We emphasized ``scalar'' to indicate
the alternative possibility that $f$ also transforms in some matrix
representation of the Lorentz group. E.g., it may transform as a Lorentz
vector, which we indicate by a vector index
\be
\delta_\omega f_a =[\omega\cdot M,f_a]=
\frac{i}{2}\omega^{bc}\eta_{a[b}f_{c]}~,
\ee
(c.f. the transformation of $P_a$ in (\ref{poin})). If it transforms as a
spinor instead, we endow $f$ with a spinor index, and a Lorentz
transformation reads
\be
\delta_\omega f_\a=[\omega\cdot
M,f_\a]=\frac{i}{2}\omega^{bc}(\C_{bc})_\a^{~\b}f_\b~,
\ee
where the Dirac algebra is
\be
\{\C_a,\C_b\}=2\eta_{ab}\cdot \bf{1}~,
\ee
and
\be
\C_{ab}\equiv\frac{1}{2}\C_{[a}\C_{b]}~.
\ee
Combining (\ref{itran}), (\ref{lore}) and (\ref{delta}) with the
possibility of a matrix representation and defining
\be
\delta f=i[\omega\cdot M+\xi \cdot P,f]~,
\la{Mtrans}
\ee
we see that we may represent the generators of $ISO(D-1,1)$ as operators on
the fields $f$ as follows:
\bea
P_a&\to& i\partial_a\nno\\
M_{ab}&\to& \frac{i}{2}x_{[a}\partial_{b]}-i\bbM~,
\la{popal}
\eea
where $\bbM$ is the appropriate matrix representation.

\subsection{Internal symmetries}

In addition to the transformation properties under the LG described in the
previous subsection, the fields may also transform in some representation
of an \underline{internal} symmetry group $\cG$. We indicate this by an
additional index $i$ on the fields. Thus, e.g., $f_\a^i(x)$ is a spinor
field which transforms in some matrix representation of $\cG$,
\be
\delta_\l f_\a^i=\l^I(B_I)_j^if_\a^j~,
\ee
where $\l$ is a transformation parameter which is taken to depend on
$x\in\cM$ for gauge symmetry. Popular internal (gauge) symmetry groups are
$\cG=U(1)$ (electro magnetsim),
$\cG=SU(2)$ (weak interactions) and $\cG=SU(3)$ (strong interactions).

It is of course tempting, in the name of unification, to try to find a
larger group  which encompasses both the Poincar\'e group and the internal
symmetry group in a non-tivial way. All such attepts came to an halt in
the late 1960's due to the famous ``No-Go'' theorem of Coleman and Mandula
(CM)
\cite{Coleman:1967ad}, where the requirements of a relativistic quantum
theory are used to limit the possibilities. In brief (leaving out some
technicalities) it states that if
\begin{quote}
(1) the S-matrix is based on a \underline{local relativistic field theory}
in space time,\\
(2) there are only a \underline{finite number of different particles}
associated with one particle states at a given mass,\\
(3) there is an \underline{energy gap} between the vacuum and the
one-particle states,
\end{quote}
then:
\begin{quote}
The most general \underline{Lie-algebra} of symmetries of the S-matrix has
generators $P_a, M_{ab}$ and $B_I$, where the $B_I$'s are 
Lorentz scalars and belong to a compact Lie-group $\cG$.
\end{quote}
The setting for this theorem is really $D=4$, so the conclusion is that
the group structure has to be $SO(3,1)\otimes\cG$.

\subsection{Supersymmetry}

In the 1970's, with the advent of supersymmetry, it was realized that
there is a loop-hole in the CM theorem, and it was extended
by Haag, Lopuszanski and Sohnius to allow for $Z_2$ \underline{graded}
Lie-algebras \cite{Haag:1975qh}. The result may be most simply stated by
giving the most general super-algebra allowed (in $D=4$). In addition to
the Poincar\'e algebra  (\ref{poin}), we also have
\bea
&&[M_{ab},B_I]=0\nno\\ 
&&[P_{a},B_I]=0\nno\\
&&[B_I,B_J]=ic_{IJ}^{~~K}B_K\nno\\
&&\{Q^i_\a,Q_\b^j\}=2\delta^{ij}(\C^aC)_{\a\b}P_a
+C_{\a\b}Z^{[ij]}+(\C_5C)_{\a\b}Y^{[ij]}\nno\\
&&[M_{ab},Q^i_\a]=\frac{1}{2}(\C_{ab})_\a^{~\b}Q^i_\b\nno\\
&&[B_I,Q^i_\a]=(B_I)_i^{~j}Q^i_\a\nno\\
&&[P_a,Q^i_\a]=0\nno\\
&&[\cO,Z]=[\cO,Y]=0~,
\la{susy}
\eea
where the three first realations say that the generators of the group 
\footnote{Note that the internal group has to be $O(\cN)$ or a
subgroup thereof. $c_{IJ}^{~~K}$ are its structure constants.} 
$\cG =O(\cN)$ are Poincar\'e scalars, in accordance with the
CM theorem. The radically new structure is carried by the $\cN$ odd
generators $Q$. As seen from (\ref{susy}) they are translationally
invariant Lorentz spinors that carry a non-trivial representation of the
internal group $\cG$. They  come together with the anticommutator
$\{~,~\}$, under which they close to a translation plus terms that depend
on the central charges $Z$ and $Y$. (The last relation in (\ref{susy}) is meant to
indicate that $Z$ and $Y$ commute with all generators.) The central
charges are antisymmetric in their $\cG$ indices, and C, finally, is the
charge conjugation matrix.

The spinors in (\ref{susy}) are \underline{Majorana spinors}, i.e., they
obey the ``reality condition'' $Q=C\bar Q^T$. In general, in $D$
dimensions the spinors have $2^{\left[\frac{D}{2}\right]}$ complex
components (where square bracket denotes `integer part of'). Depending on
the dimension $D$, one may impose ``reality'' and/or chirality conditions
on the spinors according to the following table (adapted from
\cite{Rocek:1992sw})\footnote{The table refers to space-time signature
(-++...). In (-++....+-), which is sometimes considered, there are other
possibilities. Note also that we are discussing conditions over and above
the Dirac equation which is always assumed for the spinor
\underline{fields}.}

\bigskip

\begin{tabular}{|p{1.3in}|c|c|c|c|c|c|c|c|c|c|}
\hline
D&11&10 &9&8&7&6&5&4&3&2\\
\hline
Spinor type&M&MW&M&W&D&W&D&W&M&MW\\
Real spinor dim&32&16&16&16&16&8&8&4&2&1\\
Real/Complex&R&R&R&C&C&C&C&C&R&R\\
N&1&2&2&2&2&4&4&8&16&(16,16)\\
&&1&1&1&1&2&2&4&8&(8,8)\\
&&&&&&1&1&2&4&(4,4)\\
&&&&&&&&1&2&(2,2)\\
&&&&&&&&&1&(1,1)\\
\hline
\multicolumn{11}{c}{}\\
\multicolumn{11}{c}{Table 1}
\end{tabular}

In this table, $M$ denotes Majorana, $D$ denotes Dirac and $W$ denotes
Weyl conditions. The Majorana condition was given above, Dirac just means a
Dirac spinor, i.e., no additional constaraints, and the Weyl condition in
even $D$ dimensions is 
\be
\psi=P_-\psi=\frac{1}{2} (1-\C_{D+1})\psi ~,
\ee
where $\psi$ is called a Weyl spinor and $\C_{D+1}$ is the totally
antisymmetrized product of the $D$ Dirac-matrices (suitably normalized to
make $P_-$ a projection operator.) In four dimensions one may impose
either the Majorana or the Weyl condition, as we will see below. In two
dimensions, finally, by $(p,q)$ we have indicated the possibility of having
separate right and left moving supersymmetries.\footnote{One may in fact
introduce this possibility also, e.g., in $D=6$ and $10$.}

The reader may ask what restricts the entries in Table 1 to $D\leq 11$
and/or $N\leq 16$. The reason is as follows. We know what equations
various spins should obey, and we also know the spin content of the
irreducible representations of spersymmetry (see subsection 3.3). For $N=8$ in
$D=4$ the spin content is $(2,3/2,1,1/2,0)$, whereas higher $N$ will necessarily
contain spin $\geq 2$. But ``higher spin''  field theories (with a
finite number of higher spin fields) are in general unphysical as
interacting  theories. Since $N=8$ in $D=4$ is $N=1$ in $D=11$, this sets
the limit.

\subsection{$D=4$ Supersymmetry}

In $D=4$ we illustrate explicitly the equivalence between Majorana and
Weyl spinors as well as how to build the tensors from spinor
representations. 

In $D=4$ the Weyl projection operators are given by $P_\pm=\frac{1}{2}
(1\pm\C_5)$. We utilize the isomorphism $Spin(3,1)\approx SL(2,\bbC)$ to
introduce a convenient notation for the Weyl spinors. Let $\Psi_\a$ now be
a spinor that transforms with the SL(2,\bbC) matrix $N_\a^\b$, and denote
by $\bar\Psi^{\dot\a}$ a spinor that transforms in the  conjugate
representation according to $\bar N_{\dot\a}^{\dot\b}$. Introducing also
the two sets of SL(2,\bbC) matrices $\sigma_a=(1,\underline \sigma)$ and 
$\tilde\sigma_a=(1,-\underline \sigma)$ with $\underline\sigma$ being the
Pauli matrices, we use a representation of the Dirac algebra where
\be
\C_a=\left(\begin{array}{cc}0&\sigma_a\\
\tilde\sigma_a&0\end{array}\right)~.
\ee
The relations between $\sigma$ and $\tilde\sigma$ may be stated as
\be 
(\tilde{\sigma}_a)^{\dot\a\a}=\epsilon^{\dot\a\dot\b}\epsilon^{\a\b}
(\sigma_a)_{\b\dot\b}
~,
\ee
where
\be
(\epsilon_{\a\b})=\left(\begin{array}{cc}0&-1\\
1&0\end{array}\right)=(\epsilon_{\dot\a\dot\b})=-(\epsilon^{\a\b})=
-(\epsilon_{\dot\a\dot\b})~.
\ee
In this representation
\be
C=\left(\begin{array}{cc}\epsilon_{\a\b}&0\\
0&\epsilon^{\dot\a\dot\b}\end{array}\right)~,
\ee
and a Majorana spinor may be written
\be
\Psi=\left(\begin{array}{c}\Psi_\a\\\bar\Psi^{\dot\a}
\end{array}\right)~.
\la{maj}
\ee
It is a straightforward matter to convince one-self that (\ref{maj}) indeed
satisfies $C\bar\Psi^T=\Psi$, and that 
\bea
&&P_+\Psi=\left(\begin{array}{c}\Psi_\a\\0\end{array}\right)\\
&&P_-\Psi=\left(\begin{array}{c}0\\\bar\Psi^{\dot\a}
\end{array}\right)~,
\eea
thus explicitly demonstrating the equivalence between the Majorana and
Weyl representations in four dimensions. In fact, in four dimensions the
latter representation is often preferred. In that context it is also
 convenient to represent vector indices as pairs of spinor
indices according to
\be
V_a\to (\s ^a)_{\a\dot\a}V_a\equiv V_{\a\dot\a}~.
\ee
This notation becomes particularly useful when discussing
representations of susy in \underline{superspace}.

\section{Superspace}

The main references for this section are the books
\cite{Gates:1983nr},\cite{Bagger:1990qh},\cite{Freund:1986ws} and
\cite{Buchbinder:1998qv}, but also the review article \cite{Rocek:1992sw}.

\subsection{Induced representation}

Superspace is defined via the natural generalization of the 
Minkowski-space construction described in Sec 2.2 above. Denoting the
graded ($N=1$)  Poincar\'e group by SISO(D-1,1) and specifying to $D=4$,
the relevant part of the superalgebra reads
\bea
&&\{Q_\a,\bar Q_{\dot \a}\}=P_{\a\dot\a}\nno\\
&&[M_{\a\b},Q_\gamma]=\frac{i}{2}\epsilon_{\gamma(\a}Q_{\b)}\nno\\
&&[M_{\a\b},P_{\gamma\dot\gamma}]=
\frac{i}{2}\epsilon_{\gamma(\a}P_{\b)\dot\gamma}\nno\\
&&[M_{\a\b},M_{\gamma\d}]=\frac{i}{2}(\epsilon_{\gamma(\a}M_{\b)\d}
+\gamma\leftrightarrow\d )~,
\eea 
where the antisymmetric generator of Lorentz transformations, $M_{ab}$,
is represented by its irreducible (symmetric) spinor parts  according to
\be
M_{ab}\approx
M_{\a\dot\a\b\dot\b}=i\epsilon_{\a\b}\bar
M_{\dot\a\dot\b}+i\epsilon_{\dot\a\dot\b} M_{\a\b}~.
\ee
From the algebra we exponentiate to get the group elements. This requires
introduction of Grassmann valued (anti commuting) spinor parameters.
\footnote{This exponentiation with parameters that are nilpotent is not
mathematically well defined. For this reason, mathematicians prefer to 
extend the functions on the previously discussed quotient by allowing them
to depend on Grassmann parameters instead. Operationally, I believe the net
result amounts to the same thing. See, e.g., \cite{Deligne:1999qp} for a
stringent definintion of Superspace from this point of view.} A general group
element is thus written
\be
g=e^{i(\xi\cdot P+\epsilon\cdot Q+\omega\cdot M)}~,
\ee
where we use the short hand $\epsilon\cdot Q$ for
$\epsilon^\a Q_\a+\bar\epsilon^{\dot\a}\bar Q_{\hat\a}$, and $\omega\cdot
M$ for $\omega^{\a\b}M_{\a\b}+\bar\omega^{\dot\a\dot\b}\bar
M_{\dot\a\dot\b}$. In analogy with our discussion of Minkowski space
$\cM$, we parametrize a point in the neighbourhood of the identity in
$SISO(3,1)/SO(3,1)$ by $x$ and $\theta$ according to
\be
h(x,\theta)\equiv h(Z^A)=e^{i(x\cdot P+\theta \cdot Q)}~,
\ee
(where $Z^A=x^a, \theta^\a,\bar\theta^{\dot\a}$), and find the action on
$x$ and $\theta$ through $h(Z')=g\cdot h$ mod SO(3,1). We then represent
the generators as differential operators on \underline{superfields},
i.e., on funtions $\phi(z)$. We first state the result and then supply
the necessary explanations. The operators are (cf. eqn (\ref{popal})
\bea
&&P_{\a\dot\a}=i\frac{\partial}{\partial
x^{\a\dot\a}}\equiv i\partial_{\a\dot\a}\nno\\
&&Q_\a=i\frac{\partial}{\partial\theta^{\a}}
+\frac{1}{2}\bar\theta^{\dot\a}\partial_{\a\dot\a}
\equiv
i\partial_\a+\frac{1}{2}\bar\theta^{\dot\a}\partial_{\a\dot\a}\nno\\
&&\bar Q_{\dot\a}=i\frac{\partial}{\partial\bar\theta^{\dot\a}}
+\frac{1}{2}\theta^{\a}\partial_{\a\dot\a}
\equiv
i\bar\partial_{\dot\a}+\frac{1}{2}\theta^{\a}\partial_{\a\dot\a}\nno\\
&&M_{\a\b}=\frac{i}{2}x_{(\a}^{~\dot\gamma}\partial_{\b )\dot\gamma}
+i\theta_{(\a}\partial_{\b )}-i\bbM_{\a\b}~.
\la{supop}
\eea
Here the $\theta$'s, like all the spinors, are anticommuting, which means
that derivatives with respect to $\theta$ is defined as Berezin
integrals/deivatives \cite{Berezin:1966}. We take the derivatives to act
from the left according to $\partial_\a \theta^\b =\delta_\a^\b$, and the
corresponding integration is $\int d\theta^\a \theta^\b =\delta^\b_\a$.
There is a wealth of results on the geometry of superspace, but we shall
only need a few items. 

The \underline{covariant derivatives} are differential operators $D$ that
anti-commute with the supersymmetry generators $\{D,Q\}=0$. They are
\bea
&&D_\a=\partial_\a+\frac{i}{2}\bar\theta^{\dot\a}\partial_{\a\dot\a}\nno\\
&&\bar
D_\a=\bar\partial_{\dot\a}+\frac{i}{2}\theta^{\a}\partial_{\a\dot\a}~,
\la{covder}
\eea
and their existence might have been anticipated from the fact that left
and right multiplication commutes and that the $Q$'s were defined using
left multiplication. From the point of view of superspace geometry their
(anti) commutation relation
\be
\{D_\a,\bar D_{\dot\a}\}=i\partial_{\a\dot\a}~,
\ee
signals that even ``flat'' superspace has torsion. The usefulness of the
$D$'s lies in the fact that they anticommute with the $Q$'s, since this
allow us to impose invariant conditions on the super fields.

\subsection{Superfields}

The differential operators (\ref{supop}) allow us to represent
supersymmetry on fields oveer superspace $\cM^{(p,q)}$ as we represent the
Poincar\'e group on fields over $\cM$ (here $(p,q)$ denotes $p$ bosonic
and $q$ spinorial coordinates). For example, using the explicit form of
the $Q$'s given in (\ref{supop}) we evaluate the anticommutator of two
$Q$'s acting on a scalar superfield $\phi(z)$
\be
\{Q_\a,\bar Q_{\dot\a}\}\phi(z)=i\partial_{\a\dot\a}\phi(z)~.
\ee
From the point of view of functions on Minkowski space, a superfield is a
collection of ordinary fields over $\cM$. This is seen if we make a formal
Taylor expansion in the Grassmann coordinates $\theta$ of, e.g., a real (so
called vector) superfield
\bea
\phi(z)&=&C(x)+\theta^\a\chi_\a(x)+\bar\theta^{\dot\a}\bar\chi_{\dot\a}(x)
-\theta^2M(x)-\bar\theta^2\bar M(x)
+\theta^\a\bar\theta^{\dot\a}A_{\a\dot\a}(x)\nno\\
&&-\bar\theta^2\theta^\a\lambda_\a(x)-\theta^2\bar\theta^{\dot\a}\bar\lambda_\a(x)
+\theta^2\bar\theta^2B(x)~.
\eea 
Although the Taylor series quickly terminates in this case (N=1 in four
dimensions), it is more economical to define the component fields using
the covariant derivatives (\ref{covder}). With $|$ denoting ``the $\theta$
independent part of'', the above component fields are 
\bea
&&\phi (z)|=C(x) \qquad D_\a\phi (z)|=\chi_\a (x) \qquad \bar D_{\dot\a}(z)
|=\bar\chi_{\dot\a}(x)\nno\\
&&[D_\a,\bar D_{\dot\a}]\phi(z)|=A_{\a\dot\a}(x) \qquad D^2\phi(z) |=M(x)
\qquad \bar D^2\phi(z) |=\bar M(x)\nno\\
&&-\bar D^2D_\a\phi(z)|=\l(x) \qquad D^2\bar D_{\dot\a}\phi(z) |=\bar
\l(x)\qquad D^2\bar D^2\phi(z)|=B(x)~,
\eea
where $D^2\equiv D^\a D_\a$.
The ``supermultiplet'' of $\cM$ fields represented by $\phi(z)$ and
transforming into each other under supersymmetry transformations is thus a
collection of scalar, spinor and vector fields\footnote{In a physical
model, not all of these fields will be dynamical. They have different
mass-dimensions and some of the fields will be auxiliary.}
\be
(C,\chi, \bar\chi,A_a,M,\bar M, \l,\bar\l, B)~.
\ee 
In analogy to (\ref{Mtrans}), a supersymmetry transformation of a
superfield is
\be
\delta \phi(z)=i[\e^\a Q_\a+\bar\e^{\dot\a}\bar Q_{\dot\a},\phi]~.
\la{suftf}
\ee
To illustrate the result on the component fields, we first introduce the
concept of a \underline{chiral} superfield, which is simply a (complex)
superfield $\Phi$ which satisfies
\be
\bar D_{\dot\a}\Phi=0~.
\ee
As mentioned earlier, this is a covariant condition, i.e., $\bar
D_{\dot\a}\delta\Phi=\delta(\bar D_{\dot\a}\Phi)$, as is  seen from
(\ref{suftf}) and $\{Q,\bar D\}=0$. We define the components of $\Phi$ as
follows
\be
\Phi |=\cA(x)~, \qquad D_\a\Phi |=\l _\a(x) ~,\qquad D^2\Phi |=\cF(x)~.
\la{chicom}
\ee
With, correspondingly,
\be
\delta\Phi |=\delta\cA(x)~, \qquad D_\a\delta\Phi |=\delta\l _\a(x)
~,\qquad D^2\delta\Phi |=\delta\cF(x)~,
\ee
we find the component transformations
\be
\delta\cA =-\e^\a\l_\a~,\qquad \delta\l
_\a=\e_\a\cF-i\bar\e^{\dot\a}\partial_{\a\dot\a}\cA~,\qquad
\delta\cF=-i\bar\e^{\dot\a}\partial^\a_{\dot\a}\l_\a~.
\ee
In fact, starting from these transformations, one may show that the
algebra clooses on all the fields.

\subsection{Representations}

In the previous subsection we saw how supersymmetry may be represented on
superfields. In particular, a chiral superfield was seen to be a smaller
representation  than an arbitrary superfield (it has fewer component
fields). The question thus arises of what are the irreducible
representations of supersymmetry. 

Staying with the superfields in four dimensions, we first give the
projection operators that project onto the irreducible superfields. This
is in analogy with the way a Lorentz vector $V_a$ is split into
irreducible pieces according to
\be
V_a=[(\Pi^L+\Pi^T)V]_a~,
\ee
where 
\bea
&&(\Pi^L)_a^b\equiv\partial^{-2}\partial_a\partial^b\qquad
(\Pi^T)_a^b\equiv
\partial^{-2}\delta_a^{[b}\partial^{c]}\partial_c~,\nno\\
&&~~~~~~
\eea
(such that $\Pi^L+\Pi^T=1$). Explicitly 
\be
(\Pi^LV)_a=\del^{-2}\del_a(\del^bV_b)\equiv \del^{-2}\del_aS~,\qquad 
(\Pi^TV)_a=\del^{-2}\d^b_a\del^c\del_{[c}V_{b]}\equiv
\del^{-2}\del^cF_{ca}~,
\ee
where $S$ and $F$ are irreducible representations of the Poincar\'e group.
The corresponding operators on
superfields are (for $N=1$ in four dimensions)
\bea
&&\Pi_0=\del^{-2}D^2\bar D^2\nno\\
&&\Pi_1=-\del^{-2}D^\a\bar D^2D_\a\nno\\
&&\Pi_2=\del^{-2}\bar D^2 D^2~,
\eea
such that $\Pi_0+
\Pi_1+
\Pi_2=1$. This is not quite sufficient, $\Pi_1$ has to be further
specified as $\Pi_{1\pm}$, where
\be
\Pi_{1\pm}\psi=-\del^{-2}D^\a\bar D^2D_\a\frac{1}{2}(\psi+\bar\psi)~.
\ee
Then an arbitrary (complex) scalar superfield $\Phi$ contains a chiral
superfield, two vector superfields and an
antichiral superfield (in that order) according to
\be
\Phi=\Pi_0\Phi +\Pi_{1\pm}\frac{1}{2}(\Phi +\bar\Phi)+\Pi_2\Phi~,
\ee
thus displaying the irreducible parts of the superfield.

Another question regarding representations of supersymmetry has to do with
the particle content and representations as states in a Hilbert space. 
We will use Wigners ``little group'' method to find those.

The N-extended supersymmetry algebra
in four dimensions involves the anticommutator (see (\ref{susy}))
\be
\{Q^i_\a,Q^j_\b\}=2\d^{ij}(\C^aC)_{\a\b}P_a~.
\ee 
For particles with mass $m\ne 0$ we choose $P_a=(-m,0,0,0)$. Rescaling the
charges, $Q\to\tilde Q$, by a factor $(m)^{-\frac{1}{2}}$ we have the
Clifford algebra
\be
\{\tilde Q^i_\a,\tilde Q^j_\b\}=\d_{i\a,j\b}~,\qquad i\a=1,...,4N~.
\ee
rewriting this in Weyl notation,
\bea
\{\tilde Q^i_\a,\tilde Q^j_\b\}&=&\{\tilde{ \bar Q}^i_{\dot\a},\tilde {\bar
Q}^j_{\dot\b}\}=0\nno\\
 \{\tilde Q^i_\a,\tilde{\bar Q}^j_{\dot\b}\}&=&\d_{ij}\d_{\a\dot\b}~,
\eea
we recognize  a set of 2N pairs of  annihilation and creation  operators,
$\tilde Q^i_\a\equiv a_\a^i$ and $\tilde {\bar Q}^i_{\dot\a}\equiv
 a^{\dagger i}_{\dot\a}$. Introducing the Clifford vaccum
$|0>$ such that
$a_\a^i|0>=0$, a general state is 
\be
|n_{11},n_{12},...,n_{1N},n_{21},...,n_{2N}>=\prod_{\a=1,2~~
i=1,2,...N}(a^{\dagger i}_{\dot\a})^{n_{\a i}}|0>~,
\ee
where $n_{\a i}$ denotes the occupation number of the state created by
$a^{\dagger i}_{\dot\a}$. There are clearly $2^{2N}$ such states.
For example, when N=1 , the possibilities are
\be
|0>~, \qquad a{^\dagger }_{\dot\a}|0>~, \qquad a{^\dagger }_{\dot\a}a{^\dagger
}_{\dot\b}|0>=-\frac{1}{2}\e_{\dot\a\dot\b}a^\dagger\cdot a^\dagger |0>~,
\ee
representing a Lorentz scalar, a spinor (two states) and a scalar
respectively.

The massless representations are similarily derived. Starting from the
massless four momentum $P_a=(-P,0,0,P)$, we have the anticommutation
relation
\be
\{Q^i_\a,\bar Q^j_{\dot\b}\}=2\d_{ij}P_{\a\dot\b}~,
\ee
where
\be
P_{\a\dot\b}=2P\left(\begin{array}{cc}1&0\\0&0\end{array}\right)~.
\ee
In a way analogous to the massive case, this leads to a set of
annihilation and creation  operators, $a^i$ and $a^{\dagger i}$ which 
step down and up half a unit in
helicity and
satisfy
\be
\{a^i,a^{\dagger i}\}=\d^{ij}~.
\ee
Again we introduce a Clifford vacuum which is annihilated by $a^i$, and
create the states with $a^{\dagger i}$. Since we only have one set of
operators for each $i$, instead of two, there are $2^{N}$ states in the
massless representation.

With $j_{MAX}\equiv J$ depending on the helicity of the Clifford vacuum, we
find the following table of massless states and their multiplicity for
various number of supersymmetries N.

\bigskip

\begin{tabular}{|p{0.55in}|c|c|c|c|c|c|c|c|}
\hline
N= ~
Helicity&1&2&3&4&5&6&7&8\\
\hline
J&~1~~&1~~&1~~&1~~&1~~&1~~&1~~&1~~\\
J-$\frac{1}{2}$&1&2&3&4&5&6&7&8\\
J-1&&1&3&6&10&15&21&28\\
J-$\frac{3}{2}$&&&1&4&10&20&35&56\\
J-2&&&&1&5&15&35&70\\
J-$\frac{5}{2}$&&&&&1&6&21&56\\
J-3&&&&&&1&7&28\\
J-$\frac{7}{2}$&&&&&&&1&8\\
J-4&&&&&&&&1\\
\hline
\multicolumn{9}{c}{}\\
\multicolumn{9}{c}{Table 2}
\end{tabular}

The multiplicity is simply the binomial
coefficient 
\be
\left(\begin{array}{c} N\\ k\end{array}\right)
\ee
for the totally antisymmetric product of $k$ creation operators. One
observes from Table 2 that for N=8 we need a $J$ of at least 2, for N=4 
$J$ has to be at least 1, for the smallest (CPT conjugate) multiplet with
helicities $J,....-J$. For other N, e.g, for N=1, we cannot choose a $J$
to satisfy CPT conjugation, and for a physical theory we have to add the
charge conjugated states.

\section{Supersymmetry and Complex Geometry}

The complex geometry in this section is from the books \cite{Yano1},
\cite{Yano2}, and the background material relating to supersymmetry is found in
\cite{Alvarez-Gaume:1980zi}, \cite{Zumino:1979et},
\cite{LuisDan2}. The main part of the section, however, is from the articles
\cite{Lindstrom:1983rt},\cite{Hull:1986pq} and
\cite{Hitchin:1987ea}.

\subsection{Notation}

In this subsection we collect some definitions and notation needed later.\\
For any $d=2n$ dimensional real manifold $\cM$ with coordinates $x^i$, we
may locally introduce complex coordinates as
\be
z^i=\left\{\begin{array}{c}z^A=x^i+ix^{i+n}\\
\bar z^{\bar A}=x^i-ix^{i+n}\end{array}\right\}\qquad i=1,...,n.
\ee
A mixed second rank tensor $J^i_j$ such that
$J^i_mJ^m_j=-1$ is called an \underline{almost complex structure} on $\cM$.
A metric
$g_{ij}$ which preserves  $J^i_j$ 
\be
J^i_j g_{im}J^m_n=g_{jn}, \qquad\Rightarrow  J^i_jg_{in}\equiv
J_{jn}=-J_{nj}~,
\ee
is called an \underline{almost hermitean} metric.
To make everything globally well defined and ensure that there
exist canonical complex coordinate patches related by holomorphic
transition functions, integrability conditions are needed. They may be
phrased as the vanishing of the Nijenhuis tensor
\be
N_{ij}^{~~k}= J^n_{[i}\del_{|n|} J^k_{j]}+ J^k_n\del_{[j} J^n_{i]}=0~.
\ee
The integrability conditions remove ``almost''  from
the definitions above. In the canonical coordinates the
complex structure takes the form
\be
 J^i_j=\left(\begin{array}{cc} i\d_A^B & 0\\0&-i\d_{\bar
A}^{\bar B}\end{array}\right)~,
\ee
and the components $g_{AB}$ and $g_{\bar A\bar B}$ of the hermitean metric
vanish.\\

If we further require the fundamental 2-form
\be
\omega \equiv  J^i_jg_{ik}dx^j\wedge dx^k=2ig_{A\bar A}dz^A\wedge d{\bar
z}^{\bar A}~,
\la{funtwo}
\ee
to be closed, the manifold is \underline{K\"ahler}. In such a manifold, the
Levi-Civita covariant derivative of the complex structure vanishes
\be
\nabla_i J^k_j=0~,
\ee
and the metric has a \underline{K\"ahler potential} $K$
\be
g_{A\bar A}=\frac{\del^2 K}{\del z^A\del \bar z^{\bar A}}
\ee
The converse is also true, if $\nabla J=0$ then the Nijenhuis tensor
vanishes and $g=\del\bar\del K$.

When the manifold carries three covariantly constant complex structures 
$J^{(X)}, X=1,2,3$, and these complex structures satisfy the $SU(2)$
algebra
\be
J^{(X)j}_iJ^{(Y)k}_j=-\d^{XY}\d_i^k+\e^{XYZ}J^{(Z)k}_i~,
\ee
the geometry is \underline{hyperk\"ahler}.

The various spaces described in this section are also characterized by
their \underline{holonomy}. The holonomy group $\cH_p$ at a point $p\in\cM$
is the subgroup of the tangent space group obtained by
paralell transporting vectors around closed loops in $\cM$. The restriction
to contractible loops is the \underline{restricted} holonomy $\cH_{p'}$.
When
$\cM$ is simply connected $\cH_p\approx\cH_{p'}$ and is always a subgroup
of $GL(d,\bbR)$. When $\C$ is the Levi-Civita connection, the holonomy
group is further a subgroup of $O(d)$, so a subgroup of $O(2n)$ for a
complex manifold. For a K\"ahler manifold it is smaller: E.g., if $\cM$ is
Ricci-flat the holonomy is $\subset SU(n)$.

\subsection{Nonlinear sigma models}
A link between supersymmetry and complex geometry was first established
in the context of supersymmetric \underline{non-linear sigma models},
(NLSM's),
\cite{Zumino:1979et}, \cite{LuisDan2}, \cite{Alvarez-Gaume:1980zi}. A
sigma model is a map from a manifold $\cM$, oftent taken to be
space-time, and a Target space $\cT$
\be
\Phi^A:\cM\to \cT~, 
\ee
mapping the coordinates
\be
x^a\in\cM\to \cT\ni\Phi^A(x)~.
\ee
This map is obtained by extremizing the action
\be
S=\int dx~ G_{AB}(\Phi)~\del_a\Phi^A\del_b\Phi^B\eta^{ab}~,
\la{nlsm}
\ee
which gives the equation
\be
\eta^{ab}\del_a\Phi^B\nabla_B\del_b\Phi^A=0~,
\ee
with 
\be
\nabla_AV^B=\del_AV^B+\Gamma_{AC}^{~~A}V^C~,
\ee
the target space covariant derivative. The Levi-Civita connection $\Gamma$
is formed from the target space metric $G_{AB}$,
\be
\Gamma_{AC}^{~~A}\equiv \frac{1}{2}G^{CD}(G_{D(A,B)}-G_{AB,C})~.
\ee

To get an indication of how the relation between NLSM's and complex
geometry we look at an example:\\

The complex projective space
\be
\bbC\bbP^{(n)}=U(n+1)/U(n)\times U(1)~,
\ee
is K\"ahler. Now $U(n+1)/U(n)\times U(1)$ may be thought of as the surface
\be
u^A\in \bbC^{n+1};\quad \sum_{I=1}^{n+1}\bar u^{\bar A}u^A=1~,
\la{quot}
\ee
in $\bbC^{n+1}$. The space $\bbC\bbP^{(n)}$ is thus given by the
equivalence class\footnote{There are only phase independence left from the
projective requirement due to the constraint in (\ref{quot}).}
\be
(u^1,...,u^{n+1}) \approx e^{i\phi}(u^1,...,u^{n+1})~,
\la{phase}
\ee
with $u^A$ as in (\ref{quot}). How may we describe this in terms of a NLSM?
We want the model to incorporate the structure of the manifold and to also
provide us with a metric on that manifold. If we promote the
coordinates in (\ref{quot}) to functions from some space $\cM$ and take
the sigma model action to be
\be
S=\int dx \del_a\bar u^{\bar A}(x)\del^au^A(x);\qquad \sum_{I=1}^{n+1}\bar
u^{\bar A}(x)u^A(x)=1~,
\ee
we have a start. But we still have to encode the independence of phases
(\ref{phase}) at each point. In physics terms this is the question of how to
promote the rigid $U(1)$ symmetry to a local one, to \underline{gauge} a sigma
model.  It entails
introducing a gauge field $A_a(x)$ via minimal coupling
\be
S\to S_G=\int dx (\del_a +iA_a(x))\bar u^{\bar
A}(x)(\del^a-iA_a(x))u^A(x)~.
\la{ga}
\ee
Next we eliminate $A$ by extremizing $S_G$. This does not break the gauge
invariance, it means that we choose a particular $A$ expressed in the
other fields. In terms of that particular $A$ 
\be
S_G\to \int dx \left(\del_a \bar u^{\bar A}\del^au^A+\frac{1}{4}
(\bar u^{\bar A}\stackrel{\leftrightarrow}{\del}_au^A)
(\bar u^{\bar B}\stackrel{\leftrightarrow}{\del}^au^B)\right)~.
\ee
Finally, we rewrite this in coordinates that solve the constraint in
(\ref{quot})
\bea
&&u^a=\frac{1}{\sqrt{1+z\cdot\bar z}}z^A,~, \qquad A=1,...,n \nno\\
&&u^{n+1}=\frac{1}{\sqrt{1+z\cdot\bar z}}~,
\eea
where $z\cdot\bar z\equiv z^A\bar z^{\bar A}~, A=1,...,n$.
This gives
\be
S_G=\int dx \frac{1}{1+z\cdot\bar z} \left(\d^{AB}-\frac{z^A\bar z^{\bar
B}}{1+z\cdot\bar z}\right)\del_a\bar z^{\bar A} \del^az^B~.
\la{fstud}
\ee
We recognize the Fubini-Study metric on $\bbC\bbP^{(n)}$. The corresponding
K\"ahler potential is $K=ln(1+z\cdot\bar z)$.

This example is our first encounter with a \underline{quotient
construction}, which will play an important role later.

\subsection{The bosonic quotient construction}

In the example in the previous subsection, we start from the K\"ahler
manifold
$\bbC^{n+1}$ with potential $K=\bar u^Au^A$ and construct another,
$\bbC\bbP^{(n)}$ as the space of gauge orbits by ``gauging the isometries''
and choosing a particular gauge potential that extremizes the
action.\footnote{We are glossing over the presense of the
constraint in (\ref{quot}) to bring out the essentials.}These ideas
generalize. 

Suppose we have a NLSM (\ref{nlsm}) with target space $\cT$:
\be
S=\int dx~ G_{AB}(\Phi)~\del_a\Phi^A\del_b\Phi^B\eta^{ab}~,
\la{nlsm2}
\ee
and an \underline{isometry}, i.e., a vector field $k(\Phi)$ such that
\be
\d\Phi^A=\l^qk^B_q(\Phi)=[\l^qk^B_q(\Phi)\del/\del
\Phi^B,\Phi^A]=\cL_{\l\cdot k}\Phi^A~,\qquad \cL_{\l\cdot k}G_{AB}=0~,
\la{isom}
\ee
where $\cL_{\l\cdot k}$ denotes Lie derivative along $k$.
Since the variation of the action (\ref{nlsm2}) is
\be
\d S=\int dx\left(\del_a\Phi^A\del_b\Phi^B\eta^{ab}\cL_{\l\cdot
k}G_{AB}\right)~,
\ee
it follows that an isometry is an invariance of the action $S$. In a
general situation, the isometry will be non-abelian
\be
[k_q,k_p]=c_{qp}^{~~r}k_r~,
\la{kalg}
\ee
corresponing to a non-abelian isometry group $\cG$.

We gauge the isometry (\ref{isom}) by substituting
\be
\del_a\Phi^A\to\nabla_a\Phi^A\equiv
\del_a\Phi^A-A_a^qk_q^A=\del_a\Phi^A-[A_a^qk_q^B\del/\del\Phi^B,\Phi^A]~.
\ee
(C.f. (\ref{ga})). This results in
\be
S\to S_G=\int dx~ G_{AB}(\Phi)~\nabla_a\Phi^A\nabla_b\Phi^B\eta^{ab}~.
\la{gnls}
\ee
Proceeding as in the example, we eliminate $A$ by extremising $S_G$, which
gives
\be
A_a^q=\bbH^{-1pq}G_{AB}k^A_p\del_a\Phi^B~,
\ee
where
\be
\bbH_{qp}\equiv k_q^AG_{AB}k_p^B~,
\ee
and
\be
S_G=\int
dx\left(G_{AB}-\bbH^{-1pq}k_{pA}k_{qB}\right)\del_a\Phi^A\del^a\Phi^B
\equiv
\int dx\tilde G_{AB}\del_a\Phi^A\del^a\Phi^B~.
\ee
It is easy to see that the new target space metric $\tilde G_{AB}$
projects onto the original manifold modulo the $k$-orbits, i.e., it is a
metric on the quotient space $\cT /\cG$.

The quotient construction described above gives a new target space
geometry (the coset) starting from one which has isometries. It remains to
see under which conditions the two geometries are of the same typer (as in
the $\bbC^{n+1}\to\bbC\bbP^{(n)}$ example). In particular we shall be
interested in when the supersymmetry of the original NLSM is carried over
to the quotient. To that end we first need to study supersymmetic NLSM's.

\subsection{N=1 Supersymmetric nonlinear sigma models}

We want to find a supersymmetric extension of (\ref{nlsm}), and
first need to introduce actions in superspace. Schematically, an action is
written
\be
S=\int dx d\th \cL(\Phi, D_\a\Phi,...)~,
\ee
where the measure $dx d\th$ will depend on the kind of superspace under
consideration. Explicitly, for $\cM^{(p,q)}$ the full superspace measure is
$d^pxd^q\theta$. (We will mostly consider $(p,q)=(4,4)$.) As introduced in
subsection 3.1, the integral over $\th$ is equivalent to a derivative,
$\int d\th=\del/\del\th =D_\th|$, a fact that we will often use. E.g., the
invariance of this action under a susy transformation may be shown as
follows (recall that $|$ denotes ``the
$\th$-independent part of'')
\bea
\d S&=&\int dx d\th \d\cL=\int dx d\th \e\cdot Q\cL=\int dx D^q \e\cdot
Q\cL|\nno\\
&&=\int dx\e\cdot Q D^q \cL|=\int dx\e\cdot D D^q \cL|\doteq 0~,
\eea
where the last relation $\doteq 0$ means  ``$=0$ up to total
derivatives'', and follows from the supersymmetry algebra for a
product of more than $q~D$'s.

Specializing to four dimensions, the superfields $\Phi$
that contain scalar component fields but not vectors are the chiral
superfields
$\bar D_{\dot \a}\Phi=D_\a\bar\Phi=0$. A general action in superspace for 
a set $\Phi^A$ of such fields is
\be
S=\int d^4xd^2\th d^2\bar\th K(\Phi^A,\bar \Phi^{\bar A})~,
\la{snlsm}
\ee
where we use Weyl spinor notation. Keeping the definition (\ref{chicom})
in subsection3.2 in mind, we expand the action,  again exploring the relation
between
$\th$-derivatives and integrals
\bea
&&\int d^4xd^2\th d^2\bar\th K(\Phi^A,\bar \Phi^{\bar A})\nno\\
&&=\int d^4xD^2 \bar D^2 K|=\int d^4x\left( - 2K_{A\bar B}\del A^A\del\bar
A^{\bar A}+...\right)~
\la{ssnlsm}
\eea
where indices on $K$ denote derivatives with respect to $\Phi$'s and where only
the purely bosonic content is displayed in the last equaility. This shows that
as a bosonic sigma model, we are dealing with a K\"ahler target space with
K\"ahler potential
$K$ and metric 
$K_{A\bar B}$ (ignoring the $-2$, which is due to our conventions).
In keeping with this observation, we rename the lowest component in
(\ref{chicom}),
$A^A\to z^A$ (and $\l\to\psi$), and interpret the remaining terms in the
expansion geometrically. We find
\bea
\int d^4xD^2 \bar D^2 K|=\int d^4x&&\left\{-2G_{A\bar B}\left(\del z^A\del
\bar z^{\bar B}-i\psi^A\del\bar \psi^{\bar B}-i\del\psi^A\bar \psi^{\bar
B}-\cF^A\bar\cF^{\bar B}\right)\right.\nno\\
&&+\C_{\bar{AB}C}(\cF^C\bar\psi^{\bar B}\cdot
\bar\psi^{\bar A}-2i\psi^C\del\bar z^{\bar B}\bar\psi^{\bar A})\nno\\
&&+\C_{AB\bar C}(\bar \cF^{\bar C}\psi^B\cdot
\psi^A+2i\psi^A\del z^B\bar\psi^{\bar C})\nno\\
&&\left. +(R_{C\bar AD \bar B}+G^{E\bar
E}\C_{CD\bar E}\C_{\bar{BD}E})\psi^D\cdot\psi^C\bar\psi^{\bar
B}\cdot\bar\psi^{\bar A}\right\}~,
\eea
where we have introduced the curvature $R_{C\bar AD \bar B}$ and
connection $\C_{\bar{AB}C},\C_{AB\bar C}$ for the metric $G_{A\bar B}$.
The spinorial contractions are indicated by `$\cdot$' or are the obvious
ones. Eliminating the auxiliary field $\cF$ finally gives the
fully geometric form of the component action
\be
\int d^4x\left\{-2G_{A\bar B}\left(\del z^A\del
\bar z^{\bar B}-i\psi^A\cD\bar \psi^{\bar B}-i\cD\psi^A\bar \psi^{\bar
B}\right)
+R_{C\bar AD \bar B}\psi^D\cdot\psi^C\bar\psi^{\bar
B}\cdot\bar\psi^{\bar A}\right\}~,
\ee
where the $\C \del z \psi\bar\psi$ terms are incorporated in the
derivative terms $\psi\cD\bar \psi$.

To conclude this section, a supersymmetric
NLSM in four dimensions  necessarily has a K\"ahler target space $\cT$. The
canonical complex coordinates are the lowest components of the chiral
superfields. Integrability et c is thus manifest (locally).

\subsection{Isometries in K\"ahler spaces}

In complex coordinates, the isometry (\ref{isom}) becomes
\bea
\d z^A&=&\l^qk_q^A\nno\\
\d \bar z^{\bar A}&=&\l^q\bar k_q^{\bar A}~.
\la{holism}
\eea
For a holomorphic isometry $k=k(\Phi)$ and $\bar k=\bar k(\bar\Phi)$, and
the requirement that they leave $S$ in (\ref{snlsm}) invariant means that
they have to leave $K$ invariant up to a \underline{K\"ahler gauge
transformation}
\be 
\d K(\Phi,\bar\Phi)=\l^q(K_Ak^A_q(\Phi)+K_{\bar A}\bar k_q^{\bar
A}(\bar\Phi))=\l^q(\eta_q(\Phi)+\bar\eta_q(\bar\Phi))~.
\la{kgtf}
\ee
The right hand side of (\ref{kgtf}) will give zero in the superspace
integral due to the (anti)chirality of the fields. In fact, the condition
(\ref{kgtf}) for the holomorphic $k$'s is sufficient to show that 
$\cL_k\del\bar \del K=0$, i.e., that they generate an isometry and satisfy
Killing's equation
\be
\nabla_Ak_{\bar A}+\nabla _{\bar A}\bar k_A=0~.
\ee
The relation (\ref{kgtf}) only determines  $\eta$ up to an imaginary
constant. This is reflected in an ambiguity in the 
the (real)
\underline{Killing potential} $X_q(\Phi,\bar \Phi)$ defined by
\bea
&&k^A_qK_A=iX_q+\eta_q\nno\\
&&\bar k^{\bar A}_qK_{\bar A}=-iX_q+\bar\eta_q~.
\la{killpot}
\eea
Clearly $X_q$ is correspondingly defined only up to a real constant. From
(\ref{killpot}) it follows, using the properties of $k$, that
\bea
&&k_{q\bar B}\equiv k^A_qK_{A\bar B}=iX_{q\bar B}\nno\\
&&\bar k_{qB}=-iX_{qB}~,
\eea
hence the name ``Killing potential''. It further follows that 
\be 
\bar k^{\bar B}_qX_{p\bar B}+k^B_pX_{qB}=0~,
\ee
and, hence, that
\be
\d X_p=i\l^q(\bar k_{[q}^{\bar A}X_{p]\bar A}+k_{[q}^AX_{p]A})~.
\ee
This expression for the transformation of the Killing potential will be
needed later.

For holomorphic Killing vectors, the algebra (\ref{kalg}) becomes\
\be
k^A_{[p}k^B_{q]},_A=c_{pq}^{~~Êr}k^B_r~,
\la{delK}
\ee
and its complecx conjugate. In conjunction with the transformation of the
K\"ahler potential $K$
\be
\d K=\l^q(K_ Ak^A_q+K_{\bar A}\bar k^{\bar A}_q)=\l^q(\eta_q+\bar\eta_q)~,
\ee
and analyticity, we may use (\ref{delK}) to derive the transformations of
$\eta$:
\bea
&&k^A_{[p}\eta_{q]A}=c_{pq}^{~~Êr}\eta_r+io_{pq}\nno\\
&&\bar k^{\bar A}_{[p}\bar \eta_{q]\bar A}=c_{pq}^{~~Êr}\eta_r-io_{pq}~.
\la{obstr}
\eea
It is important for the gauging of the isometries that the $\eta$'s transform
equivariantly. When the constants $o_{pq}$ are not removable, they thus
constitute an \underline{obstruction} to gauging the isometries. 
From the Jacobi identities one finds that they have to satisfy
$o_{p[q}c_{rs]}^{~~t}=0$. This is the case if
$o_{pq}=c_{pq}^{~~r}\xi_r$ for some real constant $\xi$, and the shift
$\eta\to\eta+i\xi$ then removes the obstructions, except for invariant
abelian subgroups. Indeed, for semi-simple groups, even non-compact ones, we may
choose
$\xi_q=c_{qp}^{~~r}o_{rs}g^{ps}$, with $g$ the Killing metric.

As an illustration of the previous discussion, let us look at the an
example where the isometry group is Abelian and the obstructions not
removable. Take  the K\"ahler potential to be $K=\Phi\bar\Phi$
corresponding to the flat metric $G=\bf{1}$. Then the translations
generated by
\bea
&&k_1=\del /\del\Phi+\del /\del\bar\Phi\equiv \del+\bar\del\nno\\
&&k_2=i(\del-\bar\del)~,
\eea
are isometries.
From the variation $\d K$, we find
\be
\eta_1=\Phi~, \qquad \eta_2=-i\Phi~.
\ee
Calculating the effect of a transformation as in (\ref{obstr}), we have
\be
k_{[1}^A\eta_{2]A}=-2i \quad \Rightarrow o_{12}=-2i~.
\ee
Since the isomety is abelian, this obstruction is not removable, and the
implication is that we can only gauge a linear combination of $k_1$ and
$k_2$.

\subsection{Gauging isometries in N=1 susy sigma models}

let us first discuss some generalities before taking on the isometries.
We study chiral fields that transform under some representation of a
Yang-Mills group $\cG$
\be
\Phi ^A{}'=\left(e^{i\L}\right)^A_B\Phi^B \qquad ~, \bar\Phi
^{\bar A}{}'=\bar\Phi^{\bar B}\left(e^{-i\bar\L}\right)^{\bar A}_{\bar B}~,
\la{lingag}
\ee
where $\L^A_B\equiv\L^q(T_q)^A_B$ with $\L^q(x,\th,\bar\th)$ a chiral
superfield and
$T_q$ the generators of the lie-algebra $\gg$ of $\cG$
\be
[T_q,T_p]=ic_{pq}^{~~r}T_r~.
\ee
Since $\L\ne\bar\L$, the group $\cG$ acts on $\Phi$ and $\bar\Phi$ through
its complexification 
$\cG^C$ ($\Rightarrow T_q\to (T_q,iT_q)$).

The gauge potential is an adjoint vector superfield $V=V^qT_q$. It
transforms as
\be
e^{V'}=\left(e^{i\bar\L}\right)e^V\left(e^{-i\L}\right)~,
\la{tilde}
\ee
which means that we may define a superfield $\tilde\Phi^A$ from $\bar
\Phi^A$ which transforms in $\cG_\L$ rather than in $\cG_{\bar\L}$:
\be
\tilde \Phi\equiv \bar\Phi^B\left(e^{V}\right)^A_B~, \qquad \Rightarrow
\tilde\Phi{}'=\left(\tilde\Phi e^{-i\bar\L}\right)~.
\la{fitwid}
\ee
This is precisely what is needed to construct invariant actions. To this
end we also introduce the gauge covariant superspace derivatives
\be
\nabla_{\cA}=(\nabla_\a,\nabla_{\dot\a},\nabla_{\a\dot\a})
\equiv(e^{-V}D_\a e^V,\bar D_{\dot\a},-\{\nabla_\a,\nabla_{\dot\a}\})~.
\ee
With these tools we write the gagued NLSM action in superspace as (cf
(\ref{ssnlsm}))
\be
\int d^4xd^2\th d^2\bar\th K(\Phi^A,\bar \Phi^{\bar A})
\to \int d^4xd^2\th d^2\bar\th K(\Phi^A,\tilde \Phi^{\bar A})~,
\ee
and instead of (\ref{chicom}) we use an the component expansion 
\be
z^A=\Phi^A|~,\qquad \psi_\a=\nabla_\a\Phi^A|~, \qquad
\cF^A=\nabla^2\Phi^A|~,
\ee
for the chiral fields along with
\be
A^q_{\a\dot\a}=i(\nabla_{\a\dot\a}|-\del_{\a\dot\a})^q~,\qquad
\l^q_\a=i\bar D^2\nabla_\a |\equiv W_\a|~,\qquad
D^q{}'=-\frac{i}{2}\{\nabla^\a,W_\a\}|~,
\la{veccom}
\ee
for the physical components of the vector multiplet.

Invariance of the ungauged action (\ref{ssnlsm}) under an isomorphism was
shown above to be equivalent to $\d K= \l(\eta+\bar\eta)$. Now that we are
promoting the constant $\l$ to a superfield $\L$ (and $\bar \L$), this is
no-longer necessarily true. E.g.,
\be
\int d^2\th d^2\bar\th \l^q\bar\eta_q=0~,
\ee 
but
\be
\int d^2\th d^2\bar\th \L^q\tilde\eta_q\ne 0~,
\ee
in general. To amend this, we introduce a new chiral superfield $\z$ with
transformation properties (in the ungauged case)
\be
\d \z=\eta_q(\Phi^A)\l^q~,\qquad \d \bar\z =\bar \eta_q(\bar\Phi^{\bar
A})\l^q~.
\ee
If the isometry under consideration is
generated by $k$, we now define new holomorphic Killing vectors $k'$ in the
enlarged target space with coordinates $\Phi,\bar\Phi,\z,\bar\z$:
\be
k'_q\equiv k^A_q\del_A+\eta_q\del_\z~,\qquad \bar k'_q\equiv
\bar k^{\bar A}_q\bar \del_{\bar A}+\bar\eta_q\bar\del_{\bar\z}
\la{newkill}
\ee
The symmetries generated by $k'$ leave the new K\"ahler potential
$K'\equiv K(\Phi,\bar\Phi)-\z-\bar\z$ {\em invariant}. The
new action is independent of  $\z$ but it has important consequences for the
gauged. It is this K\"ahler potential we will use when gaguging the
isometries.

The transformations (\ref{lingag}) describe transformations linearly
realized on $\Phi$ and does not cover general isometries in arbitrary
coordinates. To cover the  general case we must gauge the isometries
acting as in (\ref{isom}), or (\ref{isom}) for the holomorphic $k$'s: 
\be
 \d \Phi^A =\L^qk^A_q, \qquad \d \bar \Phi^{\bar A}=\bar\L^q\bar k^{\bar
A}_q~.
\la{gend}
\ee
The appropriate generalization of (\ref{tilde}) is
\be
\tilde\Phi^A\equiv e^{\cL_{iV\cdot \bar k}}\bar \Phi~,
\ee
i.e., the action of an exponentiated Lie-derivative along the direction
$iV^q\bar k_q$, representing a finite gauge transformation with parameter
$V^q$. Accordingly, for the case at hand with Killing vectors $k$' as in
(\ref{newkill})
\bea
\tilde\z &=&e^{\cL_{iV\cdot \bar k'}}\bar\z\nno\\
&&=\left(1+\left({e^{\cL'}-1}\over{\cL'}\right)\cL'\right)\bar \z=\bar
\z+i\left({e^{\cL'}-1}\over{\cL'}\right)\bar\eta_qV^q\nno\\
&&=\bar \z+i\left({e^{\cL}-1}\over{\cL}\right)\bar\eta_qV^q~,
\eea
where $\cL{}'\equiv \cL_{iV\cdot \bar k'}$,
 and the prime is removed in the last equality because $V$ and $\eta$
are independent of $\z$. As noted earlier, the $\bar\z$ term,
is irrelevant in the action, and thus the gauged action is
\bea
S&=&\int d^4x d^2\th d^2\bar\th \left(K(\Phi,\tilde\Phi)-\z-\tilde
\z\right)\nno\\
&&=\int d^4x d^2\th d^2\bar\th
\left(K(\Phi,\tilde\Phi)-i\left({e^{\cL}-1}\over{\cL}\right)\bar\eta_qV^q
\right)~.
\eea
Finally, we use the relation (\ref{killpot}) to eliminate $\eta$ in
favour of the Killing potential $X$ (which also entails removing the
tilde from $\Phi$ in $K$) 
\be
\int d^4x d^2\th d^2\bar\th
\left(K(\Phi,\bar\Phi)+\left({e^{\cL}-1}\over{\cL}\right)X_qV^q
\right)~.
\la{nisoneg}
\ee
The last term is hermitean, although not manifestly so. Through the 
ambiguity in the definition of $X$ there is the possibility  
to include a so called Fayet-Iliopoulos term for each $U(1)$ factor in
$\cG$, i.e., a term of the type $c_qV^q$. (See the discussion of
obstructions in subsection 4.5 above). The action (\ref{nisoneg}) solves
the problem of gauging isometries of
$N=1$ supersymmetric NLSM's.

We close this subsection with a simple example of a K\"ahler quotient
construction.\\ Let us again look at the $\bbC\bbP^{(n)}$ model
discussed in subsection 4.2, but now from the point of view of
superspace. We start from the flat space K\"ahler potential in
$\bbC^{(n+1)}$ which is 
\be
K(\Phi,\bar\Phi)=\sum_{A=1}^{n+1}\Phi^A\bar\Phi^{\bar
A}=e^{(\phi^0+\bar\phi^0)}\left(1+\sum_{a=1}^n\phi^a\bar\phi^{\bar
a}\right)~,
\ee
where the last equality involves an obvious field redefinition and displays
one of the isometries of the model. The corresponding Killing vector is
\be
\del\phi^A=\l^qk^A_q\del_A=i\l^q\d^0_q\del_0~,\qquad
\del\bar\phi^{\bar A}=\l^q\bar k^{\bar
A}_q\bar\del_{\bar A}=-i\l^q\d^0_q\bar\del_0~.
\ee
We gauge this isometry by letting
$\phi^0+\bar\phi^0\to\phi^0+\bar\phi^0+V$, which amounts to introducing
$\tilde\Phi^0$ in this case. We may also use the freedom discussed after
(\ref{nisoneg}) to include a Fayet-Iliopoulos term. Hence
\be
K(\Phi,\bar\Phi)\to
K(\Phi,\tilde\Phi)-cV=e^{(\phi^0+\bar\phi^0)}\left(1+\sum_{a=1}^n\phi^a\bar\phi^{\bar
a}\right)~.
\ee
We find the new quotient K\"ahler potential by extremizing the
corresponding action with respect to  $V$:
\be
\d V \Rightarrow V= -ln(1+\sum \phi^a\bar \phi^{\bar a})-
\phi^0-\bar\phi^0+ const.¬,
\ee
which gives the quotient K\"ahler potential
\be
K'(\phi,\bar\phi)=c~ln(1+\sum \phi^a\bar \phi^{\bar a})~,
\ee
and again we recognize the $\bbC\bbP^{(n)}$ K\"ahler  potential for the
metric  (\ref{fstud}).

\subsection{N=2 supersymmetric nonlinear sigma models}

To formulate the $N=2$ supersymmetric NLSM's we ideally 
want a
$N=2$ superspace where both supersymmetries are manifest. This means that
we need to introduce a second set of $\th$'s and extend the integration
measure accordingly, so that an action will be written as
\be
S=\int d^4x d^4\th d^4\bar\th {\cL}~.
\ee
However, such an action cannot accomodate a Lagrangian
$\cL(\Phi_H,\bar\Phi_H)$, where $\Phi_H$ is the smallest $N=2$
representation, a so called \underline{hypermultiplet} corresponding  a
pair of
 $N=1$ chiral superfields. To be more precise, a
dimensional analysis of the measure shows that such an action will not
have the right bosonic content for a NLSM. There are ways around this.
Enlarging the superspace by additional bosonic coordinates one may find
invariant subspaces and corresponding subintegrals that give correct
results. We do not discuss these {\em projective} supererspaces
\cite{Gates:1984nk}-\cite{Gonzalez-Rey:1998qh} and {\em harmonic} superspaces
\cite{Galperin:2001uw} here, though. Instead our
discussion of
$N=2$ NLSM's will be entirely in terms of $N=1$ superfields in
$N=1$ superspace. Our starting point will thus be the action (\ref{ssnlsm})
\be
\int d^4xd^2\th d^2\bar\th K(\Phi^A,\bar \Phi^{\bar A})~.
\la{nistwoa}
\ee
One supersymmetry is thus manifest due to the $N=1$ formalism. For the
second supersymmetry we make the ansatz
\be
\d \Phi^A=\bar D^2(\bar\e\bar\O^A)~, \qquad \d \bar\Phi^{\bar A}=
D^2(\e\O^{\bar A})~.
\la{second}
\ee
 The reason for the covariant derivatives is for
the second supersymmetry to commute with the first, and they come squared
to preserve the (anti-) chirality of the fields. Here
$\O=\O(\Phi,\bar\Phi)$, so they represent the general situation. The
requirements of closure of the supersymmetry algebra and invariance of
the action will constrain the $\O$'s and reveal an interesting target
space geometry.

The superfield transformation parameter satisfies
\be
\bar D_{\dot \a}\e=D^2\e=\del_{\a\dot\a}\e =0~. 
\ee
Closure of the
non-manifest supersymmetry means that 
\be
[\d _{\e^1},\d _{\e^2}]\Phi^A=i(\bar D^{\dot\a}\bar
\e^2D^\a\e^1-(1\leftrightarrow 2))\del_{\a\dot\a}\Phi^A~,
\ee  
and implies
\be
\bar\O^A_{\bar B}\O^{\bar B}_B=-\d^A_B~, \qquad \bar\O^C_{[\bar
D}\bar\O^A_{\bar B]}=0~,
\la{close}
\ee
along with their hermitean conjugate.(An
additional condition turns out to be a field equation of the model.) 
Additional subscript again represent
derivatives with respect to the fields $\Phi$ and $\bar\Phi$. 
Invariance of the action (\ref{nistwoa}) implies the further constraints
\bea
&&K_{A\bar B}\bar\O^A_{\bar C}=-K_{A\bar C}\bar\O^A_{\bar B}\nno\\
&&K_{A\bar B}\bar\O^A_{\bar C\bar D}+K_{A\bar C\bar D}\bar\O^A_{\bar
B}=0\nno\\ &&K_{A\bar B}\bar\O^A_{\bar C D}+K_{A\bar B D}\bar\O^A_{\bar
C}=0~.
\la{inv}     
\eea
Putting all this together, we conclude that there are two additional
integrable complex structures (from (\ref{close})) which are covariantly
constant with respect to the hermitean K\"ahler metric $G=\del\bar\del K$
(from (\ref{inv}).  There are thus three covariantly constant complex
structures.  The (lowest components of the) chiral superfields are the
canonical coordinates for one of them. The complex structures are
\be
J^{(3)i}_j=\left(\begin{array}{cc} i\d_{AB} &0\\0&-i\d_{\bar
A\bar B}\end{array}\right)~\quad J^{(1)i}_j=\left(\begin{array}{cc}
0&\O^{\bar A}_B \\ \bar\O^A_{\bar B}&0\end{array}\right)~,\quad
J^{(2)i}_j=\left(\begin{array}{cc}
0&i\O^{\bar A}_B \\ -i\bar\O^A_{\bar B}&0\end{array}\right)~.
\ee
We conclude that the symmetries and invariances of a $N=2$ susy NLSM requires the
target space $\cT$ to be hyperk\"ahler.

\subsection{Isometries in hyperk\"ahler spaces}

The isometries we will consider in the $N=2$ case are \underline{tri
holomorphic}, i.e., whereas a holomorphic Killing vector preserves the
fundamental two-form (\ref{funtwo}) $\o= 2iK_{A\bar A}dz^A\wedge d\bar
z^{\bar A}$ corresponding to $J^{(3)}$, a tri-holomorphic Killing vector
preserves in addition the two-forms related to $J^{(1)}$ and $J^{(2)}$,
which means that
\be
\r_{\bar B[\bar C}\nabla_{\bar D]}\bar k^{\bar B}\equiv K_{A\bar
B}\bar\O^A_{[\bar C}\nabla_{\bar D]}\bar k^{\bar B}=0~.
\ee
This defines $\r$ (and $\bar \r$ through the hermitean conjugate
relation). Such a Killing vector has a Killing potential with respect to
each $J$, or, in arbitrary coordinates,
$k^iJ_{ij}^{(X)}=-X^{(X)},_j$. We combine $X^{(1)}$ and $X^{(2)}$ to  a
holomorphic potential $P$ and an antiholomorphic potential $\bar P$ with
respect to
$J^{(2)}\pm iJ^{(2)}$, respectively,
\be
k^A\r _{AB}=-P,_B~,\qquad \bar k^{\bar A}\bar \r_{\bar A\bar B}=-\bar
P,_{\bar B}~.
\ee
These ingredients are all needed to describe the gauging of isometries of
$N=2$ NLSM's.

\subsection{Gauging isometries in N=2 susy sigma models}

When the $N=2$ model (\ref{nistwoa}) has triholomorphic isometries they
may be gauged in a manner that closely follows the description in
subsection 4.6. The new features to do with $NÅ2$ supersymmetry is that
the scalar superfields now come in pairs of  chiral
$N=1$ field that together constitute a $N=2$ hyper multiplet. Also the
vector superfield $V^q$ gets an $N=2$ partner, a chiral 
superfield
$S^q$ .
\be
\Phi^A\to (\Phi^A_+,\Phi^A_-)~,\qquad V^q\to (V^q,S^q)~.
\ee
The second supersymmetry (\ref{second}) is affected by the gauging in
that
$\O(\Phi,\bar\Phi)\to \O(\Phi,\tilde\Phi)$. In addition, for the $N=2$ vector
multiplet it reads reads
\be
\d e^V=\e\bar Se^V+e^VS\bar\e~,\d S=-iW^\a D_\a\e~.
\ee
(See (\ref{veccom})). The gauge transformations with parameter $\L$ are
as in (\ref{gend}) with the additional
\be
\d S=i[\L,S]~.
\ee
The gauged action, invariant under the local isometries, is the
generalization of (\ref{nisoneg})
\bea
&&\int d^4x d^2\th d^2\bar\th
\left[K(\Phi,\bar\Phi)+\left({e^{\cL}-1}\over{\cL}\right)X_qV^q
+g_{pg}S^p\left(e^{-V}\bar Se^V\right)^q\right]\nno\\
&&+\left\{\int d^4xd^2\th (iS^qP_q) +h.c. \right\}~.
\la{gnit}
\eea
The possibility to add Fayet-Iliopoulos terms discussed for $N=1$
generalizes to $N=2$
\be
S_{FI}=\int d^4x d^2\th d^2\bar\th c_qV^q +\left\{\int d^4x d^2\th i\hat
c_qS^q+h.c.\right\}~,
\ee
again with a sum over abelian factors.

The action (\ref{gnit}) is the starting point for the $N=2$ quotient, the
hyperk\"ahler quotient construction \cite{Hitchin:1987ea}, and we end with an
example of this \cite{Lindstrom:1983rt}.

Starting from the action
\bea
&&\int d^4x d^2\th d^2\bar\th
\left(\bar\Phi^{\bar A}_+\Phi^A_+e^V+\Phi^A_-\bar\Phi^{\bar
A}_-e^{-V}-cV\right)\nno\\
&&+\left\{\int d^4xd^2\th ( \Phi^A_-\Phi^A_+-b)S) +h.c. \right\}~,
\la{acac}
\eea
which is a gauged flat ($\bbC^{2(n+1)}$) $N=2$ action with Fayet-Iliopoulos terms ($c$ and
$b$), invariant under the gauged abelian isometries
\be
\Phi_\pm'=e^{\pm i\L}\Phi_\pm~,\qquad
\bar \Phi_\pm'=e^{\mp\bar\L}\bar\Phi_\pm~,\qquad V'=V+i(\bar\L-\L)~.
\ee
We extremise this action with respect to the $N=2$ vector multiplet, i.e.,
with respect to $V$ and $S$
\bea
&&\d V \Rightarrow \bar\Phi^{\bar A}_+\Phi^A_+e^V-\Phi^A_-\bar\Phi^{\bar
A}_-e^{-V}=c\nno\\
&&\d S \Rightarrow \Phi^A_-\Phi^A_+=b~,
\la{mop}
\eea
where the last relation is known as the ``moment map''. With the gauge choice
\be
\Phi^{n+1}_+=\Phi^{n+1}_-\equiv \phi~,
\ee
and the redefinitions
\be
\Phi^a_\pm\equiv U^a_\pm\phi~,\qquad a=1,...,n ~,
\ee
these moment map constraints are solved. Further defining
\be
\qquad M_\pm\equiv\bar\Phi^A_\pm\Phi^A_\pm~,
\ee
we  solve the $V$ equations in (\ref{mop}) and rewrite the action
(\ref{acac}) as
\be
\int d^4x d^2\th
d^2\bar\th\left[\sqrt{c^2+4M_+M_-}-c\left(ln(c+\sqrt{c^2+4M_+M_-}~)-lnM_+\right)\right]~.
\ee
The Lagrangian density inside the square brackets is the new K\"ahler potential on the
$N=2$ quotient, which also has a hyperk\"ahler target space $\cT$. The
quotient K\"ahler potential is a generalization of that of the
$\bbC\bbP^n$ models and yields  the Calabi metrics.
\bigskip

{\bf Acknowledgement}: I gratefully acknowledge all my collaborators in
Supersymmetry. In particular Nigel Hitchin, Chris Hull, Anders Karlhede and
Martin Ro\v cek. The work is supported in parts by EU contract HPNR-CT-2000-0122
and by VR grant 650-1998368.


\begin{thebibliography}{99}


%\cite{Gates:1983nr}
\bibitem{Gates:1983nr}
S.~J.~Gates, M.~T.~Grisaru, M.~Rocek and W.~Siegel,\\
\underline{\em Superspace, 
Or One Thousand And One Lessons In
Supersymmetry}, \\Front.\ Phys.\  {\bf 58}, 1 (1983)
[arXiv:hep-th/0108200].
%%CITATION = HEP-TH 0108200;%%

%\cite{Bagger:1990qh}
\bibitem{Bagger:1990qh}
J.~Bagger and J.~Wess,
`\underline{\em Supersymmetry And Supergravity},
JHU-TIPAC-9009,  Princeton Univ. Press.

%\cite{Freund:1986ws}
\bibitem{Freund:1986ws}
P.~G.~Freund,
\underline{\em Introduction To Supersymmetry},\\
{\it  Cambridge, Uk: Univ. Pr. ( 1986) 152 P.\\ 
( Cambridge Monographs On Mathematical Physics)}.

%\cite{Buchbinder:1998qv}
\bibitem{Buchbinder:1998qv}
I.~L.~Buchbinder and S.~M.~Kuzenko,
``Ideas and methods of supersymmetry and supergravity: Or a walk through 
superspace,'' {\it  Bristol, UK: IOP (1998) 656 p}.

%\cite{Freed:1999mn}
\bibitem{Freed:1999mn}
D.~S.~Freed,
``Five lectures on supersymmetry,''
{\it  Providence, USA: AMS (1999) 119 p}.

%\cite{Deligne:1999qp}
\bibitem{Deligne:1999qp}
P.~.~Deligne {\it et al.},
``Quantum fields and strings: A course for mathematicians.  Vol. 1, 2,''
{\it  Providence, USA: AMS (1999) 1-1501}.

%\cite{Lindstrom:1983rt}
\bibitem{Lindstrom:1983rt}
U.~Lindstrom and M.~Rocek,
``Scalar Tensor Duality And N=1, N=2 Nonlinear Sigma Models,''
Nucl.\ Phys.\ B {\bf 222}, 285 (1983).
%%CITATION = NUPHA,B222,285;%%

%\cite{Hull:1986pq}
\bibitem{Hull:1986pq}
C.~M.~Hull, A.~Karlhede, U.~Lindstrom and M.~Rocek,
``Nonlinear Sigma Models And Their Gauging In And Out Of
Superspace,'' Nucl.\ Phys.\ B {\bf 266}, 1 (1986).
%%CITATION = NUPHA,B266,1;%%

%\cite{Hitchin:1987ea}
\bibitem{Hitchin:1987ea}
N.~J.~Hitchin, A.~Karlhede, U.~Lindstrom and M.~Rocek,
``hyperkahler Metrics And Supersymmetry,''
Commun.\ Math.\ Phys.\  {\bf 108}, 535 (1987).
%%CITATION = CMPHA,108,535;%%

%\cite{Peskin:1995ev}
\bibitem{Peskin:1995ev}
M.~E.~Peskin and D.~V.~Schroeder,
``An Introduction to quantum field theory,''
{\it  Reading, USA: Addison-Wesley (1995) 842 p}.

%\cite{Weinberg:1995mt}
\bibitem{Weinberg:1995mt}
S.~Weinberg,
``The Quantum theory of fields. Vol. 1: Foundations,''
{\it  Cambridge, UK: Univ. Pr. (1995) 609 p}.

%\cite{Fayet:1977cr}
\bibitem{Fayet:1977cr}
P.~Fayet and S.~Ferrara,
``Supersymmetry,''
Phys.\ Rept.\  {\bf 32}, 249 (1977).
%%CITATION = PRPLC,32,249;%%

%\cite{Salam:1978ib}
\bibitem{Salam:1978ib}
A.~Salam and J.~Strathdee,
``Supersymmetry And Superfields,''
Fortsch.\ Phys.\  {\bf 26}, 57 (1978).
%%CITATION = FPYKA,26,57;%%

%\cite{Corwin:1975fi}
\bibitem{Corwin:1975fi}
L.~Corwin, Y.~Ne'eman and S.~Sternberg,
``Graded Lie Algebras In Mathematics And Physics (Bose-Fermi
Symmetry),'' Rev.\ Mod.\ Phys.\  {\bf 47}, 573 (1975).
%%CITATION = RMPHA,47,573;%%

%\cite{Pais:1975hu}
\bibitem{Pais:1975hu}
A.~Pais and V.~Rittenberg,
``Semisimple Graded Lie Algebras,''
J.\ Math.\ Phys.\  {\bf 16}, 2062 (1975)
[Erratum-ibid.\  {\bf 17}, 598 (1975)].
%%CITATION = JMAPA,16,2062;%%


%\cite{Rittenberg:1978eg}
\bibitem{Rittenberg:1978eg}
V.~Rittenberg and M.~Scheunert,
``Elementary Construction Of Graded Lie Groups,''
J.\ Math.\ Phys.\  {\bf 19}, 709 (1978).
%%CITATION = JMAPA,19,709;%%

%\cite{Scheunert:1977wj}
\bibitem{Scheunert:1977wj}
M.~Scheunert, W.~Nahm and V.~Rittenberg,
``Irreducible Representations Of The Osp(2,1) And Spl(2,1)
Graded Lie Algebras,'' J.\ Math.\ Phys.\  {\bf 18}, 155 (1977).
%%CITATION = JMAPA,18,155;%%

%\cite{Scheunert:1977wi}
\bibitem{Scheunert:1977wi}
M.~Scheunert, W.~Nahm and V.~Rittenberg,
``Graded Lie Algebras: Generalization Of Hermitian
Representations,'' J.\ Math.\ Phys.\  {\bf 18}, 146 (1977).
%%CITATION = JMAPA,18,146;%%

%\cite{Scheunert:1976ug}
\bibitem{Scheunert:1976ug}
M.~Scheunert, W.~Nahm and V.~Rittenberg,
``Classification Of All Simple Graded Lie Algebras Whose Lie
Algebra Is Reductive. 2. (Construction Of The Exceptional
Algebras),'' J.\ Math.\ Phys.\  {\bf 17}, 1640 (1976).
%%CITATION = JMAPA,17,1640;%%

%\cite{Scheunert:1976uf}
\bibitem{Scheunert:1976uf}
M.~Scheunert, W.~Nahm and V.~Rittenberg,
``Classification Of All Simple Graded Lie Algebras Whose Lie
Algebra Is Reductive. 1,'' J.\ Math.\ Phys.\  {\bf 17}, 1626
(1976).
%%CITATION = JMAPA,17,1626;%%

%\cite{Nahm:1976mu}
\bibitem{Nahm:1976mu}
W.~Nahm, V.~Rittenberg and M.~Scheunert,
``The Classification Of Graded Lie Algebras,''
Phys.\ Lett.\ B {\bf 61}, 383 (1976).
%%CITATION = PHLTA,B61,383;%%

%\cite{Coleman:1967ad}
\bibitem{Coleman:1967ad}
S.~R.~Coleman and J.~Mandula,
``All Possible Symmetries Of The S Matrix,''
Phys.\ Rev.\  {\bf 159}, 1251 (1967).
%%CITATION = PHRVA,159,1251;%%

%\cite{Haag:1975qh}
\bibitem{Haag:1975qh}
R.~Haag, J.~T.~Lopuszanski and M.~Sohnius,
``All Possible Generators Of Supersymmetries Of The S Matrix,''
Nucl.\ Phys.\ B {\bf 88}, 257 (1975).
%%CITATION = NUPHA,B88,257;%%

%\cite{Rocek:1992sw}
\bibitem{Rocek:1992sw}
M.~Ro\v{c}ek,
``Introduction to supersymmetry,''
{\it  In Boulder 1992, Proceedings, 
Recent directions in particle theory 101-139}.

\bibitem{Berezin:1966}
F.~A.~Berezin, ``The method of second quantization'', Academic Press: New
York (1966).

\bibitem{Yano1}
K.~Yano, \underline{\em Differential geometry on complex and
almost complex spaces} \\(Pergamon, Oxford, 1965).

\bibitem{Yano2}
K.~Yano and M.~Kon, \underline{\em Structures of manifolds},
Series in Pure Mathematics, Vol.3
 (World Scientific, Singapore, 1984)

%\cite{Alvarez-Gaume:1980zi}
\bibitem{Alvarez-Gaume:1980zi}
L.~Alvarez-Gaume and D.~Z.~Freedman,
``A Simple Introduction To Complex Manifolds,''
ITP-SB-80-31.

%\cite{Zumino:1979et}
\bibitem{Zumino:1979et}
B.~Zumino,
``Supersymmetry And Kahler Manifolds,''
Phys.\ Lett.\ B {\bf 87}, 203 (1979).
%%CITATION = PHLTA,B87,203;%%

\bibitem{LuisDan2}
L.~Alvarez-Gaume and D.~Z.~Freedman:\\ In: \underline{\em
Unification of the fundamental particle interactions},\\
Ferrara,S., Ellis,J. and van Nieuwenhuizen,P. (eds.). New York:
Plenum 1980.

%\cite{Gates:1984nk}
\bibitem{Gates:1984nk}
S.~J.~Gates, C.~M.~Hull and M.~Rocek,
``Twisted Multiplets And New Supersymmetric Nonlinear Sigma Models,''
Nucl.\ Phys.\ B {\bf 248}, 157 (1984).
%%CITATION = NUPHA,B248,157;%%

%\cite{Karlhede:1984vr}
\bibitem{Karlhede:1984vr}
A.~Karlhede, U.~Lindstrom and M.~Rocek,
``Selfinteracting Tensor Multiplets In N=2 Superspace,''
Phys.\ Lett.\ B {\bf 147}, 297 (1984).
%%CITATION = PHLTA,B147,297;%%

%\cite{Grundberg:1985xr}
\bibitem{Grundberg:1985xr}
J.~Grundberg and U.~Lindstrom,
``Actions For Linear Multiplets In Six-Dimensions,''
Class.\ Quant.\ Grav.\  {\bf 2}, L33 (1985).
%%CITATION = CQGRD,2,L33;%%

%\cite{Lindstrom:1988ks}
\bibitem{Lindstrom:1988ks}
U.~Lindstrom and M.~Rocek,
``New Hyperkahler Metrics And New Supermultiplets,''
Commun.\ Math.\ Phys.\  {\bf 115}, 21 (1988).
%%CITATION = CMPHA,115,21;%%

%\cite{Lindstrom:1990ne}
\bibitem{Lindstrom:1990ne}
U.~Lindstrom and M.~Rocek,
``N=2 Superyang-Mills Theory In Projective Superspace,''
Commun.\ Math.\ Phys.\  {\bf 128}, 191 (1990).
%%CITATION = CMPHA,128,191;%%

%\cite{Lindstrom:1994mw}
\bibitem{Lindstrom:1994mw}
U.~Lindstrom, I.~T.~Ivanov and M.~Rocek,
``New N=4 superfields and sigma models,''
Phys.\ Lett.\ B {\bf 328}, 49 (1994)
[arXiv:hep-th/9401091].
%%CITATION = HEP-TH 9401091;%%

%\cite{Lindstrom:1995eh}
\bibitem{Lindstrom:1995eh}
U.~Lindstrom, B.~B.~Kim and M.~Rocek,
``The Nonlinear multiplet revisited,''
Phys.\ Lett.\ B {\bf 342}, 99 (1995)
[arXiv:hep-th/9406062].
%%CITATION = HEP-TH 9406062;%%

%\cite{Gonzalez-Rey:1998qh}
\bibitem{Gonzalez-Rey:1998qh}
F.~Gonzalez-Rey, M.~Rocek, S.~Wiles, U.~Lindstrom and R.~von Unge,
``Feynman rules in N = 2 projective superspace. I: Massless  hypermultiplets,''
Nucl.\ Phys.\ B {\bf 516}, 426 (1998)
[arXiv:hep-th/9710250].
%%CITATION = HEP-TH 9710250;%%

%\cite{Galperin:2001uw}
\bibitem{Galperin:2001uw}
See
A.~S.~Galperin, E.~A.~Ivanov, V.~I.~Ogievetsky and E.~S.~Sokatchev,
``Harmonic superspace,'' 
{\it  Cambridge, UK: Univ. Pr. (2001) 306 p} and references therein.





\end{thebibliography}
\end{document}